\documentclass[english]{IEEEtran}
\PassOptionsToPackage{ruled,linesnumbered}{algorithm2e}
\usepackage[T1]{fontenc}
\usepackage[latin9]{inputenc}
\usepackage{refstyle}
\usepackage{units}
\usepackage{algorithm2e}
\usepackage{amsmath}
\usepackage{amsthm}
\usepackage{amssymb}
\usepackage{graphicx}

\makeatletter

\AtBeginDocument{\providecommand\thmref[1]{\ref{thm:#1}}}
\AtBeginDocument{\providecommand\appref[1]{\ref{app:#1}}}
\AtBeginDocument{\providecommand\algref[1]{\ref{alg:#1}}}
\AtBeginDocument{\providecommand\figref[1]{\ref{fig:#1}}}
\AtBeginDocument{\providecommand\subfigref[1]{\ref{subfig:#1}}}
\RS@ifundefined{subsecref}
  {\newref{subsec}{name = \RSsectxt}}
  {}
\RS@ifundefined{thmref}
  {\def\RSthmtxt{theorem~}\newref{thm}{name = \RSthmtxt}}
  {}
\RS@ifundefined{lemref}
  {\def\RSlemtxt{lemma~}\newref{lem}{name = \RSlemtxt}}
  {}

\theoremstyle{plain}
\newtheorem{thm}{\protect\theoremname}
\theoremstyle{remark}
\newtheorem{rem}[thm]{\protect\remarkname}

\newcommand\ztag[1]{%
\def\@currentlabel{#1}%
\gdef\tmp{%
\addtocounter{equation}{-1}%
\def\theequation{#1}}%
\aftergroup\aftergroup\aftergroup\aftergroup\aftergroup\aftergroup
\aftergroup\aftergroup\aftergroup\aftergroup\aftergroup\aftergroup
\aftergroup\aftergroup\aftergroup\aftergroup\aftergroup\aftergroup
\aftergroup\aftergroup\aftergroup\aftergroup\aftergroup\aftergroup
\aftergroup\aftergroup\aftergroup\aftergroup\aftergroup\aftergroup
\aftergroup
\tmp}

\usepackage{mathtools,empheq}
\usepackage{cite}
\usepackage{caption}
\usepackage{refstyle}
\usepackage[subrefformat=parens,labelformat=parens]{subfig}
\usepackage[esvectg]{overarrows}
\usepackage{pgfplots}
\usepackage{grffile}
\pgfplotsset{compat=newest}
\usetikzlibrary{plotmarks}
\usepgfplotslibrary{patchplots}
\usepackage{amsmath}
\usepackage[T1]{fontenc}
\usepackage[latin9]{inputenc}
\usepackage{pgfplots}
\usepackage{grffile}
\pgfplotsset{compat=newest}
\usetikzlibrary{plotmarks}
\usepgfplotslibrary{patchplots}
\usepackage{amsmath}
\usepackage{xcolor}
\usepackage{tikz}
\usepackage{nicefrac, xfrac}
\usepackage{cases}

\newcommand{\herm}{^{\mathsf{H}}}
\newcommand{\TX}{{\mathtt{T}}}

\newcommand{\trans}{^{\mathsf{T}}}
\usepackage{amsmath}
\DeclareMathOperator{\Tr}{Tr}

\DeclareMathOperator{\diag}{diag}

\DeclareMathOperator{\sinc}{sinc}

\allowdisplaybreaks[1]

\newref{fig}{refcmd = Fig.~\ref{#1}}
\newref{subfig}{refcmd = Fig.~\subref*{#1}}
\newref{alg}{refcmd = \textbf{Algorithm~\ref{#1}}}
\newref{app}{refcmd = Appendix~\ref{#1}}

\newref{subsec}{refcmd = Section~\ref{#1}}
\newref{sec}{refcmd = Section~\ref{#1}}
\renewcommand*{\thmref}[1]{\textbf{Theorem~\ref{thm:#1}}}

\AtBeginDocument{\providecommand\eqref[1]{(\ref{eq:#1})}}

\ifdefined\showcaptionsetup
 \PassOptionsToPackage{caption=false}{subfig}
\fi
\usepackage{subfig}
\makeatother

\usepackage{babel}
\providecommand{\remarkname}{Remark}
\providecommand{\theoremname}{Theorem}

\begin{document}
\title{A Refined Alternating Optimization for Sum Rate Maximization in SIM-Aided
Multiuser MISO Systems}
\author{Eduard E. Bahingayi, \IEEEmembership{Member, IEEE}, Shuying Lin,
Murat Uysal, \IEEEmembership{Fellow, IEEE}, Marco Di Renzo, \IEEEmembership{Fellow, IEEE},
and Le-Nam Tran, \IEEEmembership{Senior Member, IEEE} \thanks{E. E. Bahingayi and L.-N. Tran are with the School of Electrical and
Electronic Engineering, University College Dublin, Belfield, Dublin
4, D04 V1W8, Ireland (e-mail: eduard.bahingayi@ucd.ie; nam.tran@ucd.ie).}\thanks{S. Lin is with the School of Telecommunications and Information Engineering,
Nanjing University of Posts and Telecommunications, Nanjing, China
(e-mail: shu\_ying\_lin@163.com).}\thanks{Murat Uysal is with the Engineering Division, New York University Abu Dhabi (NYUAD), Abu Dhabi, UAE, 129188 (e-mail: murat.uysal@nyu.edu).}\thanks{M. Di Renzo is with Universit\'e Paris-Saclay, CNRS, CentraleSup\'elec, Laboratoire des Signaux et Syst\`emes, 3 Rue Joliot-Curie, 91192 Gif-sur-Yvette, France (e-mail: marco.di-renzo@universite-paris-saclay.fr), and also with King\textquotesingle s College London, Centre for Telecommunications Research -- Department of Engineering, WC2R 2LS London, United Kingdom (e-mail: marco.di\_renzo@kcl.ac.uk).}}
\maketitle
\begin{abstract}
Stacked intelligent metasurfaces (SIMs) have emerged as a disruptive
technology for future wireless networks. To investigate their capabilities,
we study the sum rate maximization problem in an SIM-based multiuser
(MU) multiple-input single-output (MISO) downlink system. A vast majority
of pioneer studies, if not all, address this fundamental problem using
the prevailing alternating optimization (AO) framework, where the
digital beamforming (DB) and SIM phase shifts are optimized alternately.
However, many of these approaches suffer from suboptimal performance,
quickly leading to performance saturation, when the number of SIM
layers increases assuming the \emph{fixed SIM thickness}. In this
letter, we demonstrate that significant performance gains can still
be achieved, and such saturation does not occur with the proposed
method in the considered setting. To this end, we provide practical
design guidelines to improve AO-based optimization of digital precoders
and SIM phase shifts. Specifically, we show that (i) optimizing the
SIM phase shifts first yields significant performance improvements,
compared to optimizing the DB first; and (ii) when applying projected
gradient (PG) methods, which are gradually becoming more popular to
optimize the phase shifts thanks to their scalability, we find that
using an iterative PG method achieves better performance than the
single PG step, which is commonly used in existing solutions. Based
on these customizations, the proposed method achieves a higher achievable
sum rate (ASR) of up to $\ensuremath{115.53\%}$, compared to benchmark
schemes for the scenarios under consideration.
\end{abstract}


\section{INTRODUCTION}

Stacked intelligent metasurface (SIM) is the latest development in
intelligent surface technology, offering unprecedented capability
to boost both spectral and energy efficiency, and, thus, holds the
potential to revolutionize future wireless networks \cite{an2024stacked}.
SIM-aided systems integrate multiple metasurface layers into the base
station (BS) transceiver architecture. Each layer consists of numerous
meta-atoms that act as secondary signal sources for the subsequent
layer, enabling signal propagation across layers. Unlike conventional
multiple-input multiple-output (MIMO) systems, which rely on digital
baseband processors to perform signal precoding and combining, SIM
performs these operations directly in the electromagnetic (EM) domain,
as signals propagate through the metasurface layers\cite{di2024state,hassan2024efficient,an2023stacked}.
As a result, this wave-domain processing capability significantly
reduces the reliance on costly and power-hungry radio frequency (RF)
chains and complex baseband hardware, positioning SIM as a promising
architecture for next-generation wireless networks\cite{an2024stacked,an2023stacked}.

From an architectural perspective, SIM-aided systems are broadly divided
into two categories. The first category, SIM without digital beamforming
(DB) (SIMwoDB), relies entirely on wave-based beamforming (WB) for
signal processing \cite{an2023stacked,an2023stackedmu,an2025stackedmu,lin2024stacked,papazafeiropoulos2024achievablemu,hu2024joint,shi2024joint,niu2024stacked,bahingayi2025scaling}.
In this setup, data streams are fed directly into the transmit antennas,
with each antenna transmitting an independent stream. Optimization
problems in this category primarily involve the joint design of the
transmit power allocation (PA) and WB. The second category, SIM with
DB (SIMwDB), employs a hybrid beamforming approach that integrates
both DB and WB into the transceiver design \cite{papazafeiropoulos2024achievable,park2025sim,perovic2024energy,shi2025energy,NemanjaSIM}.
This hybrid architecture provides additional degrees of freedom for
beamforming optimization, leading to improved energy efficiency and
achievable sum rate (ASR) compared to SIMwoDB \cite{perovic2024energy}.

Various studies have been carried out to explore the potential of
SIM technology under different scenarios. For example, \cite{an2023stacked,bahingayi2025scaling,NemanjaSIM,papazafeiropoulos2024achievable}
focused on SIM-aided single-user (SU) MIMO systems. Specifically,
\cite{an2023stacked} investigated the achievable rate by solving
a channel fitting problem, while \cite{papazafeiropoulos2024achievable}
extended this analysis to the SIMwDB structure. Most recently, \cite{bahingayi2025scaling}
demonstrated that the achievable rate of an SIM-aided SU-MIMO system
increases with the number of SIM layers by keeping the thickness of
the SIM fixed. Meanwhile, \cite{NemanjaSIM} explored mutual information
maximization for SIM-aided SU-MIMO system, using the cutoff rate as
an alternative performance metric. The ASR of SIM-aided multi-user
(MU) multiple-input single-output (MISO) systems was studied in \cite{lin2024stacked,an2025stackedmu,papazafeiropoulos2024achievablemu},
while energy efficiency was investigated in \cite{perovic2024energy,shi2025energy}.
Additionally, the ASR performance of SIM-aided cell-free MU-MISO systems
was explored in \cite{hu2024joint,park2025sim,shi2024joint}, whereas
\cite{niu2024stacked} considered SIM-aided MU-MISO systems integrated
with joint sensing and communication. However, despite these advances,
the field of SIM-based communications is still in its infancy. A key
challenge in the design of SIM-aided systems lies in the nonconvex
and large-scale nature of the underlying optimization problems, which
hinders the derivation of globally optimal solutions. To address this,
previous studies\cite{an2023stacked,an2024stacked,an2025stackedmu,NemanjaSIM,papazafeiropoulos2024achievable,lin2024stacked,papazafeiropoulos2024achievablemu,hu2024joint,shi2024joint,niu2024stacked,bahingayi2025scaling,perovic2024energy,shi2025energy}
adopted the alternating optimization (AO) framework, where the SIM
phase shifts and DB or PA are optimized alternately.

In this study, we investigate the potential of SIM to improve the
ASR in SIM-based MU-MISO downlink systems. Specifically, we focus
on the SIMwDB architecture, since it includes SIMwoDB as a special
case. In SIM-aided systems, both DB and SIM contribute to interference
suppression, but their relative significance within the AO framework
has not been studied in the literature. Through numerical experiments,
we find that the order and degree of refinement in optimizing SIM
phase shifts significantly affect performance due to the non-convex
nature of the problem and the sensitivity of the AO method to initial
point. To this end, we propose a refined AO framework that achieves
improved ASR by reordering and refining SIM phase shift optimization.
Our contributions are as follows:
\begin{itemize}
\item When applying the AO framework to the design of SIM-aided systems,
we first reveal that the order in which beamformers and phase shifts
are optimized significantly impacts performance. Particularly, we
numerically show that it is more beneficial to optimize the SIM phase
shifts first. While a few existing studies may have implicitly adopted
this order, it has not, to the best of our knowledge, been explicitly
recognized or studied in the context of SIM optimization. This is
noteworthy since the AO framework applied to solve non-convex problems
is known to be highly dependent on the initial point.
\item For phase shift optimization, projected gradient (PG) methods have
been widely used in existing literature. In particular, from a current
iterate of the phase shifts, a \textit{single} PG step is usually
performed to obtain the next iterate. In this context, we find that
an\textit{ iterative} PG method yields significant performance improvements.
By iterative PG, we mean that PG steps are repeated until convergence
is achieved.
\end{itemize}
\emph{Notation}: Upper and lowercase boldface letters denote matrices
and vectors, respectively. $x_{i}$ is the $i$-th entry of $\mathbf{x}$,
and $\left[\cdot\right]_{i,j}$ is the $($$i$, $j$$)$-th element
of matrix. $\mathbf{\left(\cdot\right)}^{*}$, $\mathbf{\left(\cdot\right)}\trans$,
and $\mathbf{\left(\cdot\right)}\herm$ denote the conjugate, transpose,
and Hermitian, respectively. $\Tr\{\cdot\}$ and $\left\Vert \mathbf{\cdot}\right\Vert $
denote the trace and Euclidean norm, respectively. $\diag(\mathbf{\cdot})$
forms a diagonal matrix. $\mathbf{\nabla}_{\mathbf{X}}f(\cdot)$ is
the gradient of $f$ with respect to (w.r.t) $\mathbf{X}^{\ast}$.
$\mathbf{I}_{N}$ is the $N\times N$ identity matrix and $\otimes$
denotes the Kronecker product. $\Re\{\cdot\}$, $\arg\{\cdot\}$,
and $\bigl|\cdot\bigr|$ denote the real part, the angle, and the
absolute value of a complex number, respectively. $\mathbb{C}$ stands
for the complex numbers.

\section{System Model}

We consider an SIM-based MU-MISO downlink system with the SIMwDB architecture,
where a BS has $N_{t}$ antennas and $L$ SIM layers with $N$ meta-atoms
each, serving $K$ single-antenna user equipments (UEs).

Define $\theta_{n}^{l}=e^{j\psi_{n}^{l}}$, where $\psi_{n}^{l}\in[0,2\pi)$,
$\forall n\in\mathcal{N}\triangleq\{1,\dots,N\}$, and $\forall l\in\mathcal{L}\triangleq\{1,\dots,L\}$,
denote the transmission coefficient of the $n$-th meta-atom in the
$l$-th metasurface layer. The transmission coefficient vector of
the $l$-th metasurface layer is $\boldsymbol{\theta}^{l}=\left[\theta_{1}^{l},\dots,\theta_{N}^{l}\right]\trans\in\mathbb{C}^{N\times1}$,
and the corresponding diagonal matrix is $\boldsymbol{\Theta}^{l}=\diag(\boldsymbol{\theta}^{l})\in\mathbb{C}^{N\times N}$.
Furthermore, let $\mathbf{F}^{l}\in\mathbb{C}^{N\times N},\forall l\neq1,l\in\mathcal{L}$,
denote the channel response matrix between two adjacent metasurface
layers, while the matrix between the BS antenna array and the first
metasurface layer is $\mathbf{F}^{1}\in\mathbb{C}^{N\times N_{t}}$.
Based on Rayleigh--Sommerfeld diffraction theory \cite{Lin_2018},
the $(n,n')$-th entry of $\mathbf{F}^{l},\forall l\in\mathcal{L}$
is given by 
\begin{equation}
[\mathbf{F}^{l}]_{n,n'}=\frac{A\cos\psi_{n,n'}^{l}}{d_{n,n'}}\Bigl(\frac{1}{2\pi d_{n,n'}}-\frac{j}{\lambda}\Bigr)\exp\Bigl(\frac{j2\pi d_{n,n'}}{\lambda}\Bigr),\label{eq: Rayleigh-Sommerfield}
\end{equation}
for $\forall l\in\mathcal{L}$, where $A$ denotes the surface area of each meta-atom, $\lambda$ is the wavelength of the carrier frequency, $d_{n,n'}$ represents the transmission distance between the $n'$-th meta-atom of the $(l-1)$-th layer and the $n$-th meta-atom of the $l$-th layer, and $\psi_{n,n'}^{l}$ is the angle between the propagation direction and the normal to the $(l-1)$-th metasurface layer. Hence, the WB matrix is given by 
\begin{align}
\mathbf{G}= & \boldsymbol{\Theta}^{L}\mathbf{F}^{L}\boldsymbol{\Theta}^{L-1}\mathbf{F}^{L-1}\cdots\boldsymbol{\Theta}^{1}\mathbf{F}^{1}\in\mathbb{C}^{N\times N_{t}}.\label{eq:gen_A}
\end{align}

Assuming perfect channel state information and quasi-static flat-fading, the received signal at the $k$-th UE is given by 
\begin{align}
y_{k} & =\mathbf{h}_{k}\herm\mathbf{G}\mathbf{w}_{k}x_{k}+\sum\nolimits_{j\neq k}^{K}\mathbf{h}_{k}\herm\mathbf{G}\mathbf{w}_{j}x_{j}+z_{k},\label{eq:rece_signal}
\end{align}
where $\mathbf{h}_{k}\in\mathbb{C}^{N\times1}$ denotes the channel from the meta-atoms in the last SIM layer to the $k$-th UE, $\mathbf{w}_{k}\in\mathbb{C}^{N_{t}\times1}$ is the DB vector, $x_{k}$ is the transmitted symbol intended for the $k$-th UE, generated such that $\mathbf{\mathbb{E}}\bigl\{ x_{k}x_{k}^{\ast}\bigr\}=1$, and $z_{k}\thicksim\mathcal{CN}(0,\sigma^{2})$ represents additive white Gaussian noise (AWGN), where $\sigma^{2}$ is the noise power. Based on (\ref{eq:rece_signal}), the signal-to-interference-plus-noise-ratio (SINR) of the $k$-th UE is given by
\begin{align*}
\gamma_{k} & =\frac{|\mathbf{h}_{k}\herm\mathbf{G}\mathbf{w}_{k}|^{2}}{\sum_{j\neq k}^{K}|\mathbf{h}_{k}\herm\mathbf{G}\mathbf{w}_{j}|^{2}+\sigma^{2}}.
\end{align*}

\section{Problem Formulation and Proposed Solution}

In this paper, the optimization problem is formulated as 
\begin{subequations}
\label{eq:WSR}\begin{IEEEeqnarray}{rCl}
&\max_{ \boldsymbol{\theta}, \mathbf{W}}  &\quad R(\boldsymbol{\theta}, \mathbf{W})\triangleq \sum\nolimits_{k=1}^{K}\ln\left(1+\gamma_{k}\right),
\\*
({\mathcal{P}_1}) : \smash{\left\{
\IEEEstrut[9\jot]
\right.}
&{\rm s.t.}  &\quad \boldsymbol{\theta}\in\mathcal{Q}, \label{eq:modulus}
\\*
&  & \quad\sum\nolimits_{k=1}^{K}\Vert\mathbf{w}_{k}\Vert^{2}\leq P_{{\rm T}}\label{eq:SPC},
\end{IEEEeqnarray}
\end{subequations}
where $\mathbf{W}\triangleq[\mathbf{w}_{1},\dots,\mathbf{w}_{K}]\in\mathbb{C}^{N_{t}\times K}$,
$\boldsymbol{\theta}=[(\boldsymbol{\theta}^{1})\trans,\dots,(\boldsymbol{\theta}^{L})\trans]\trans\in\mathbb{C}^{NL\times1}$, $\mathcal{Q}\triangleq\{\boldsymbol{\theta}\Bigl||\theta_{n}^{l}|=1,\thinspace n\in\mathcal{N},l\in\mathcal{L}\}$, and $P_{{\rm T}}$ is the total transmit power at the BS. The constraints in (\ref{eq:modulus}) and (\ref{eq:SPC}) arise from the unit modulus of the reflection coefficient of the SIM meta-atoms and the total
transmit power budget at the BS, respectively.

We note that $(\mathcal{P}_{1})$ is a highly non-convex optimization problem due to the unit-modulus constraint of the element coefficients of the SIM meta-atoms, making a globally optimal solution difficult to obtain. To tackle this, we adopt the AO method, which is widely used for similar problems in related works. Specifically, we decouple $(\mathcal{P}_{1})$ into two subproblems for optimizing $\boldsymbol{\theta}$ and $\mathbf{W}$ alternately until convergence. In the proposed method, we optimize $\boldsymbol{\theta}$ first, which is numerically shown to achieve a significantly higher ASR. As emphasized earlier, while this ordering may have been used implicitly in prior works, this is the first work to explicitly showcase the impact of optimization order on ASR performance in SIM-aided systems.

\subsection{SIM Phase Shift Optimization}

The optimization subproblem for $\boldsymbol{\theta}$, given a fixed $\mathbf{W}$ in $(\mathcal{P}_{1})$, can be formulated as  
\begin{equation}
    (\mathcal{P}_{2}) \triangleq \{\underset{\boldsymbol{\theta}}{\max}\ R(\boldsymbol{\theta})\ |\  (\ref{eq:modulus}) \}.
\end{equation} By abuse of notation, we write $R(\boldsymbol{\theta})$ instead of $R(\boldsymbol{\theta},\mathbf{W})$, where $\mathbf{W}$ is omitted as it is a fixed variable. We note that $R(\boldsymbol{\theta})$ is continuously differentiable and the elements of $\boldsymbol{\theta}$ lie on the unit circle in the complex plane. This constraint structure makes the PG method a natural choice for solving $(\mathcal{P}_{2})$, which has led to its widespread adoption in the literature. The PG method handles the non-convex unit-modulus constraint by iteratively projecting the gradient updates onto the feasible set. Specifically, the PG step at the $m$-th iteration is given by
\begin{align}
\boldsymbol{\theta}^{(m+1)} & =\Pi_{\mathcal{Q}}\bigl(\boldsymbol{\theta}^{(m)}+\alpha^{(m)}\nabla_{\boldsymbol{\theta}}R(\boldsymbol{\theta}^{(m)})\bigr),\label{eq:GP}
\end{align}
where $\nabla_{\boldsymbol{\theta}}R(\cdot)$ denotes the complex-valued
gradient of $R(\boldsymbol{\theta})$, $\alpha^{(m)}>0$ is the step
size, and $\Pi_{\mathcal{Q}}\left(.\right)$ denotes the projection
of the argument onto the set $\mathcal{Q}$. In particular, for $\ensuremath{\mathbf{\mathbf{a}}\in\mathbb{C}^{NL\times1}}$,
the projection $\ensuremath{\mathbf{\bar{\mathbf{a}}}}=\Pi_{\mathcal{Q}}\left(\mathbf{a}\right)$
is defined element-wise as \begin{equation}
\bar{a}_i = 
\begin{cases}
\frac{a_i}{|a_i|}, & \text{if } a_i \neq 0, \\
e^{j\phi}, & \text{if } a_i = 0,\ \phi \in [0, 2\pi).
\end{cases}
\end{equation} The gradient $\nabla_{\boldsymbol{\theta}}R(\boldsymbol{\theta})$
in (\ref{eq:GP}) is given in \thmref{grad:theta}.
\begin{thm}
\label{thm:grad:theta}A closed-form expression for \textup{$\nabla_{\boldsymbol{\theta}}R(\boldsymbol{\theta})$
}is given by\textup{
\begin{equation}
\nabla_{\boldsymbol{\theta}}R(\boldsymbol{\theta})=\left[(\nabla_{\boldsymbol{\theta}^{1}}R(\boldsymbol{\theta}))\trans,\cdots,(\nabla_{\boldsymbol{\theta}^{L}}R(\boldsymbol{\theta}))\trans\right]\trans\in\mathbb{C}^{NL\times1},\label{eq:comp_gradtheta}
\end{equation}
}where $\nabla_{\boldsymbol{\theta}^{l}}R(\boldsymbol{\theta})$ is
given by
\begin{align}
\ensuremath{\nabla_{\boldsymbol{\theta}^{l}}R(\boldsymbol{\theta})} & =\sum_{k=1}^{K}\Biggl(\frac{\sum_{j=1}^{K}\mathbf{e}_{k,j}^{l\ast}\bigl(\mathbf{e}_{k,j}^{l}\bigr)\trans\boldsymbol{\theta}^{l}}{\sum_{j=1}^{K}|\bigl(\boldsymbol{\theta}^{l}\bigr)\trans\mathbf{e}_{k,j}^{l}|^{2}+\sigma^{2}}\nonumber \\
 & \qquad-\frac{\sum_{j\neq k}^{K}\mathbf{e}_{k,j}^{l\ast}\bigl(\mathbf{e}_{k,j}^{l}\bigr)\trans\boldsymbol{\theta}^{l}}{\sum_{j\neq1}^{K}|\bigl(\boldsymbol{\theta}^{l}\bigr)\trans\mathbf{e}_{k,j}^{l}|^{2}+\sigma^{2}}\Biggr),\forall l\in\mathcal{L},\label{eq:phi_derivative}
\end{align}
and $\mathbf{e}_{k,j}^{l}$ is defined in \appref{proof_of_grad_theta}.
\end{thm}
\begin{IEEEproof}
See \appref{proof_of_grad_theta}.
\end{IEEEproof}
The step size is selected via a backtracking line search, by selecting
$\alpha^{(m)}=\beta^{i}\alpha_{0},$ for the smallest $i\in\mathbb{N}$
such that
\begin{align*}
R(\boldsymbol{\theta}^{(m+1)}) & \geq\bar{R}(\boldsymbol{\theta}^{(m)},\boldsymbol{\theta}^{(m+1)})\\
 & \triangleq R(\boldsymbol{\theta}^{(m)})+\eta\bigl\Vert\boldsymbol{\theta}^{(m+1)}-\boldsymbol{\theta}^{(m)}\bigr\Vert^{2},
\end{align*}
where $\beta\in(0,1)$ and $\eta>0$ is a small constant.

\subsection{Digital Beamforming Optimization}

By fixing $\boldsymbol{\theta}$ in $(\mathcal{P}_{1})$, the optimization
problem for $\mathbf{W}$ is given by  \begin{equation}
    (\mathcal{P}_{3}) \triangleq \{\underset{\mathbf{W}}{\max}\ R(\mathbf{W})\ |\  (\ref{eq:SPC}) \}.
	\end{equation}Similarly, by abuse of notation, we use $R(\mathbf{W})$ instead of
$R(\boldsymbol{\theta},\mathbf{W})$, where $\boldsymbol{\theta}$
is omitted as it is a fixed variable. We note that problem $(\mathcal{P}_{3})$
is a classical sum-rate maximization problem for the MU-MISO downlink,
which is typically solved using the well-known weighted minimum mean
squared error (WMMSE) approach \cite{shi2011iteratively}. However,
due to the high computational complexity of WMMSE, we adopt a recently
proposed low-complexity successive convex approximation (SCA)-based
method \cite{bahingayi2024joint} for more efficient optimization.
Notably, both WMMSE and SCA achieve the same sum-rate performance,
making SCA a more practical alternative.

Furthermore, we remark that, for the SIMwoDB architecture, $\mathbf{W}$
in $(\mathcal{P}_{3})$ is a diagonal matrix with its diagonal containing
the square roots of the per-antenna power allocation values, i.e.,
$\mathbf{W}=\diag\bigl(\sqrt{p_{1}},\dots,\sqrt{p_{K}}\bigr)$, where
$p_{k}$ is the power allocated to the $k$-th transmit antenna. Most
importantly, we employ the same SCA-based method proposed in \cite{bahingayi2024joint}
to solve the corresponding power allocation problem.

The overall proposed method is outlined in \algref{AO}. 
\begin{algorithm}[t]
\SetAlgoLined
\DontPrintSemicolon
\SetKwRepeat{Do}{do}{while}
\SetKwInput{Initialize}{Initialize}
\SetKwInOut{Input}{Input}
\SetKwInOut{Output}{Output}\KwIn{ $\{\mathbf{h}_{k}\}_{k=1}^{K}$, $\{\mathbf{F}^{l}\}_{l=1}^{L}$,
and $\sigma^{2}$.}

\Initialize{ $\boldsymbol{\theta}^{(0)}$, $\mathbf{W}^{(0)}$, $\alpha_{0}>0$,
$\beta\in(0,1)$, and $\eta>0$.}

Set $q\leftarrow0$

\Repeat{$R(\mathbf{W}^{(q+1)},\boldsymbol{\theta}^{(q+1)})-R(\mathbf{W}^{(q)},\boldsymbol{\theta}^{(q)})\leq\epsilon$
}{


\Initialize{ $\boldsymbol{\theta}^{(0)}\leftarrow\boldsymbol{\theta}^{(q)}$,
$\alpha^{(0)}=\alpha_{0}$, and $m\leftarrow0$.}

\Repeat{$R(\boldsymbol{\theta}^{(m+1)})-R(\boldsymbol{\theta}^{(m)})\leq\epsilon_{\boldsymbol{\theta}}$}{

Compute $\nabla_{\boldsymbol{\theta}^{(m)}}R(\boldsymbol{\theta}^{(m)})$
using (\ref{eq:comp_gradtheta}) .\;

\Repeat{$R(\boldsymbol{\theta}^{(m+1)})\geq\bar{R}(\boldsymbol{\theta}^{(m)},\boldsymbol{\theta}^{(m+1)})$}{

$\boldsymbol{\theta}^{(m+1)}=\Pi_{\mathcal{Q}}\bigl(\boldsymbol{\theta}^{(m)}+\alpha^{(m)}\nabla_{\boldsymbol{\theta}^{(m)}}R(\boldsymbol{\theta}^{(m)})\bigr)$

\If{$R(\boldsymbol{\theta}^{(m+1)})<\bar{R}(\boldsymbol{\theta}^{(m)},\boldsymbol{\theta}^{(m+1)})$}{

$\alpha^{(m)}\leftarrow\beta\alpha^{(m)}$\;

} 					 		

}

$m\leftarrow m+1$\;

}

$\boldsymbol{\theta}^{(q+1)}\leftarrow\boldsymbol{\theta}^{(m)}$	


Compute $\mathbf{W}^{(q+1)}$ using the SCA method\cite{bahingayi2024joint}

$q\leftarrow q+1$\;

}	

\caption{AO procedure for optimizing $\boldsymbol{\theta}$ and $\mathbf{W}$.\label{alg:AO}}
\end{algorithm}
The novelty of the proposed method is further explained in the following
two remarks. It should be noted that to achieve improved ASR performance
gain requires both of the following design strategies to be applied
in tandem.
\begin{rem}[The order of optimization between $\boldsymbol{\theta}$ and $\mathbf{W}$]
 Most of the existing studies using AO optimize $\mathbf{W}$ first
and then $\boldsymbol{\theta}$ \cite{an2023stacked,NemanjaSIM,shi2025energy,an2024stacked,hu2024joint,niu2024stacked,lin2024stacked,papazafeiropoulos2024achievable,shi2024joint,an2025stackedmu}.
\emph{In the proposed method, we instead emphasize optimizing $\boldsymbol{\theta}$
first}. As demonstrated numerically in the next section, this seemingly
minor difference creates a major impact on the ASR. It is, however,
not surprising for non-convex problems, as the performance of AO depends
on the initial point. However, this effect has not been previously
reported in the literature. Our results indicate that SIM contributes
more significantly to the ASR than DB, owing to the interference suppression
capability of SIM-aided WB, which improves with the number of layers
and often surpasses that of conventional DB\cite{an2024stacked,shi2025energy,bahingayi2025scaling}.
As a result, optimizing\emph{ $\boldsymbol{\theta}$ }first enables
the system to exploit this capability more effectively, thereby accelerating
convergence to the optimal solution and enhancing the ASR.
\end{rem}
\begin{rem}[Single PG vs. iterative PG]
 In the proposed method, the PG step for the optimization of $\boldsymbol{\theta}$,
described in Lines $\ensuremath{4-13}$ of \algref{AO}, is repeated
until the stopping criterion in Line $13$ is satisfied, before moving
to the optimization of $\mathbf{W}$. We refer to the proposed method
as iterative PG. In the context of AO, usually only a single PG step
is performed for the optimization of $\boldsymbol{\theta}$ i.e.,
Lines $\ensuremath{6-11}$ are immediately followed by the optimization
of $\mathbf{W}$. In fact, this is the approach adopted in previous
studies \cite{an2023stacked,an2024stacked,an2025stackedmu,NemanjaSIM,papazafeiropoulos2024achievable,lin2024stacked,papazafeiropoulos2024achievablemu,hu2024joint,shi2024joint,niu2024stacked,shi2025energy}.
The numerical results presented next indicate that optimizing $\boldsymbol{\theta}$
using the iterative PG significantly enhances the ASR performance
compared to the single PG method.
\end{rem}

\section{Simulation Results\label{sec:SimResults}}

In this section, we present numerical simulation results to evaluate
the ASR performance of the SIM-aided MU-MISO downlink system by comparing
the proposed AO method with the iterative PG where $\boldsymbol{\theta}$
is updated first, with three AO benchmarks: (i) iterative PG ($\mathbf{W}$
is updated first), which optimizes $\mathbf{W}$ first and applies
the iterative PG to optimize $\boldsymbol{\theta}$; (ii) single PG
($\boldsymbol{\theta}$ is updated first), a variant of the proposed
AO method that employs a single PG step to optimize $\boldsymbol{\theta}$
\cite{papazafeiropoulos2024achievablemu}; and (iii) single PG ($\mathbf{W}$
is updated first), a widely used approach in previous works, where
$\mathbf{W}$ is optimized first, followed by a single PG step to
optimize $\boldsymbol{\theta}$ \cite{an2023stacked,papazafeiropoulos2024achievablemu,hu2024joint,NemanjaSIM,niu2024stacked,lin2024stacked,papazafeiropoulos2024achievable,shi2024joint,an2025stackedmu}.

We consider a carrier frequency of $f_{c}=28\thinspace{\rm GHz}$,
which gives $\lambda=10.7\thinspace{\rm mm}$. The BS and UEs have
heights of $15\thinspace{\rm m}$ and $1.6\thinspace{\rm m}$, respectively,
with UEs randomly scattered within a circle of radius $10\thinspace{\rm m}$
at a distance of $100\thinspace{\rm m}$ from the BS. The thickness
of SIM is $D=10\lambda$, so the spacing between adjacent metasurface
layers, as well as between the BS antennas and the first metasurface
layer, is $d_{s}=D/L$, while the area of each meta-atom is $A=\lambda^{2}/4$.
Unless specified otherwise, we use $N_{t}=K=4$, $N=49$, $P_{\TX}=30\,{\rm dBm}$,
and $\sigma^{2}=-104\,{\rm dBm}$. For the PG-based method, we set
$\alpha_{0}=1$, $\eta=10^{-7}$, $\beta=0.5$, and $\epsilon_{\boldsymbol{\theta}}=\epsilon=10^{-6}$.

The baseband equivalent channel vector from the last SIM layer to
the $k$-th UE is modeled as a correlated Rayleigh fading channel,
i.e., $\mathbf{h}_{k}\sim\mathcal{CN}(\mathbf{0},\varsigma_{k}\mathbf{R})$,
where $\varsigma_{k}$ represents the large-scale fading parameter,
given by $\varsigma_{k}[{\rm dB}]=10c_{1}\log_{10}(\nicefrac{4\pi d_{0}}{\lambda})+10c_{2}\log_{10}(\nicefrac{d_{k}}{d_{0}}),\forall k=1,\dots,K$,
where $d_{0}$ and $d_{k}$ denote the reference and link distances
in meters, respectively, while $c_{1}$ and $c_{2}$ are path loss
constants. We set $d_{0}=1\,{\rm m}$, $c_{1}=2$, and $c_{2}=3.5$.
The matrix $\mathbf{R}\in\mathbb{C}^{N\times N}$ denotes the spatial
correlation matrix of the last metasurface layer of the SIM. Its $(n,n')$-th
entry is given by $[\mathbf{R}]_{n,n'}=\sinc(\nicefrac{2d_{n,n'}}{\lambda})$
\cite{an2023stacked}, where $\sinc(x)=\sin(\pi x)/(\pi x)$. 
\begin{figure}[t]
\centering
\includegraphics{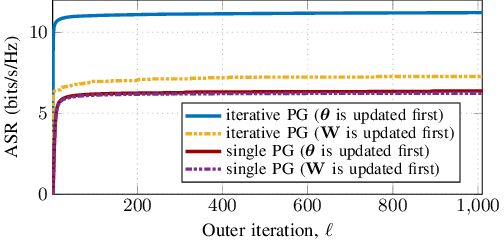}\caption{\label{fig:Convg}Convergence of the algorithms when solving $(\mathcal{P}_{1})$,
considering the SIMwDB architecture.}
\end{figure}
\begin{figure}[t]
\begin{minipage}[t]{0.5\columnwidth}%
\subfloat[\label{subfig:SIMwDB}]{\centering{}\includegraphics{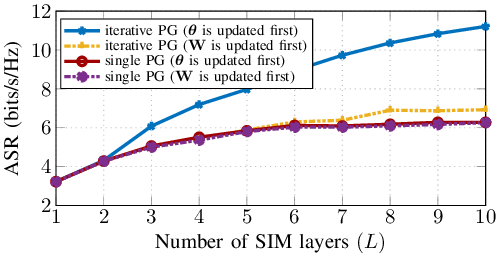}}%
\end{minipage} %
\begin{minipage}[t]{0.5\columnwidth}%
\subfloat[\label{subfig:SIMwoDB}]{\centering{}\includegraphics{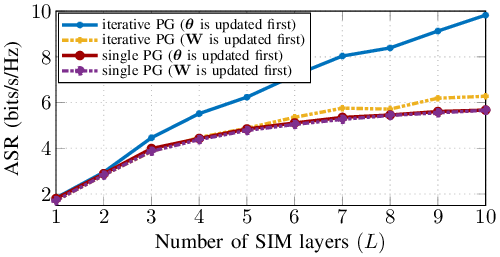}}%
\end{minipage}\caption{ASR versus the number of SIM layers for an SIM-aided MU-MISO system
with $N_{t}=K=4$, $N=49$, and $P_{{\rm T}}=30~{\rm dBm}$ (a). SIMwDB
(b). SIMwoDB.\label{fig:RvsL}}
\end{figure}

\figref{Convg} compares the convergence behavior of the considered
schemes for solving $(\mathcal{P}_{1})$, starting from the same randomly
generated initial point $\bigl\{\boldsymbol{\theta}^{(0)},\mathbf{W}^{(0)}\bigr\}$.
\figref{Convg} clearly shows that the iterative PG ($\boldsymbol{\theta}$
is updated first) scheme converges to a significantly higher ASR compared
to the other baseline schemes. On the other hand, the single PG ($\mathbf{W}$
is updated first) scheme achieves the lowest ASR. This observed performance
gap is mainly due to the higher interference suppression capability
of SIM-based WB compared to DB, which makes optimizing $\boldsymbol{\theta}$
first a more effective solution. Moreover, the iterative PG method
exploits this capability more effectively than the single PG method,
leading to an improved ASR.

\figref{RvsL} demonstrates the impact of increasing the number of
SIM layers ($L$). Both \subfigref{SIMwDB} and \subfigref{SIMwoDB}
clearly show that the ASR of the iterative PG ($\boldsymbol{\theta}$
is updated first) scheme improves significantly with $L$, whereas
the ASR of other schemes saturates for large values of $L$, which
is consistent with previous studies \cite{shi2024joint,an2025stackedmu,papazafeiropoulos2024achievable,shi2025energy,an2024stacked,an2023stacked,papazafeiropoulos2024achievablemu}.
In \subfigref{SIMwDB}, assuming $L=10$, the ASR gains of the iterative
PG ($\boldsymbol{\theta}$ is updated first) over the iterative PG
($\mathbf{W}$ is updated first), single PG ($\boldsymbol{\theta}$
is updated first), and single PG ($\mathbf{W}$ is updated first)
schemes are approximately $\ensuremath{54.77\%}$, $\ensuremath{76.79\%}$,
and $\ensuremath{78.19\%}$, respectively. Also, in \subfigref{SIMwoDB}
still assuming $L=10$, the ASR gains of the iterative PG ($\boldsymbol{\theta}$
is updated first) over the iterative PG ($\mathbf{W}$ is updated
first), single PG ($\boldsymbol{\theta}$ is updated first), and single
PG ($\mathbf{W}$ is updated first) schemes are approximately $\ensuremath{56.57\%}$,
$\ensuremath{73.21\%}$, and $\ensuremath{73.53\%}$, respectively.
These performance gains stem from the fact that SIM offers stronger
interference mitigation capability than DB in SIM-aided systems, which
is effectively exploited by the proposed method as it prioritizes
the optimization of $\boldsymbol{\theta}$ over $\mathbf{W}$, and
employs the iterative PG approach.

\begin{figure}[t]
\centering
\includegraphics{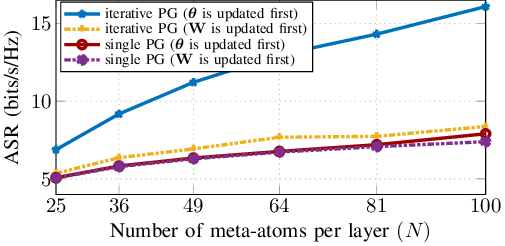}\caption{\label{fig:ASRvsN}ASR versus the number of meta-atoms per layer for
an SIM-aided MU-MISO system with $N_{t}=K=4$, $L=10$, and $P_{{\rm T}}=30~{\rm dBm}$,
considering the SIMwDB architecture.}
\end{figure}
\figref{ASRvsN} illustrates the ASR versus the number of meta-atoms
per layer $N$, assuming the number of SIM layers is fixed at $L=10$.
It is observed that the ASR of all considered schemes increases with
$N$. Notably, the iterative PG ($\boldsymbol{\theta}$ is updated
first) scheme significantly outperforms the benchmark schemes, with
performance gaps widening as $N$ increases. Specifically, at $N=100$,
the iterative PG ($\boldsymbol{\theta}$ is updated first) scheme
achieves performance gains of approximately $\ensuremath{91.35\%}$,
$\ensuremath{93.05\%}$, and $\ensuremath{115.53\%}$ over the iterative
PG ($\mathbf{W}$ is updated first), single PG ($\boldsymbol{\theta}$
is updated first), and single PG ($\mathbf{W}$ is updated first)
schemes, respectively.

\section{Conclusion\label{sec:Conclusion}}

In this paper, we have investigated the ASR performance of SIM-aided
MU-MISO downlink systems by proposing an efficient AO-based approach
to alternately optimize SIM phase shifts and DB or PA. The proposed
method prioritizes optimizing the SIM phase shifts first, followed
by DB or PA, which has been numerically shown to achieve significantly
improved ASR performance. Furthermore, when PG is adopted for phase
shift optimization, our results show that the proposed iterative PG
method yields significant ASR gains compared to the single PG step
approach commonly used in existing studies. Simulation results confirm
that the ASR attained by the proposed AO method improves by approximately
$\ensuremath{115.53\%}$ compared to benchmark schemes. These findings
provide new insights into efficient algorithm design for future SIM-based
wireless networks.\vspace{-10 pt}

\appendix[Proof of \thmref{grad:theta} \label{app:proof_of_grad_theta}]{}

First, we explicitly rewrite $(\mathcal{P}_{2})$ as function of $\boldsymbol{\theta}^{l}\thinspace\forall l\in\mathcal{L}$.
To this end, we define the following matrices: $\mathbf{G}^{L+}=\mathbf{I}_{N}$,
$\mathbf{G}^{1-}=\mathbf{F}^{1}$, $\mathbf{G}^{l-}=\mathbf{F}^{l}\boldsymbol{\Theta}^{l-1}\mathbf{F}^{l-1}\cdots\boldsymbol{\Theta}^{2}\mathbf{F}^{2}\boldsymbol{\Theta}^{1}\mathbf{F}^{1},1\leq l\leq L,$
and $\mathbf{G}^{l+}=\boldsymbol{\Theta}^{L}\mathbf{F}^{L}\boldsymbol{\Theta}^{L-1}\mathbf{F}^{L-1}\cdots\boldsymbol{\Theta}^{l+1}\mathbf{F}^{l+1},1\leq l\leq L-1.$
The gradient of $R(\boldsymbol{\theta})$ w.r.t $\boldsymbol{\theta}^{l\ast}$
is given by 
\begin{align}
\ensuremath{\nabla_{\boldsymbol{\theta}^{l}}R(\boldsymbol{\theta})} & =\sum_{k=1}^{K}\Bigl(\frac{\nabla_{\boldsymbol{\theta}^{l}}\bigl(\sum_{j=1}^{K}|\bigl(\boldsymbol{\theta}^{l}\bigr)\trans\mathbf{e}_{k,j}^{l}|^{2}+\sigma^{2}\bigr)}{\sum_{j=1}^{K}|\bigl(\boldsymbol{\theta}^{l}\bigr)\trans\mathbf{e}_{k,j}^{l}|^{2}+\sigma^{2}}\nonumber \\
 & \qquad-\frac{\nabla_{\boldsymbol{\theta}^{l}}\bigl(\sum_{j\neq k}^{K}|\bigl(\boldsymbol{\theta}^{l}\bigr)\trans\mathbf{e}_{k,j}^{l}|^{2}+\sigma^{2}\bigr)}{\sum_{j\neq1}^{K}|\bigl(\boldsymbol{\theta}^{l}\bigr)\trans\mathbf{e}_{k,j}^{l}|^{2}+\sigma^{2}}\Bigr)\label{eq:gradftheta_l}
\end{align}
where $\mathbf{e}_{k,j}^{l}=\diag\bigl(\mathbf{h}_{k}\herm\mathbf{G}^{l+}\bigr)\mathbf{G}^{l-}\mathbf{w}_{j}\in\mathbb{C}^{N\times1}$.
Thus, we obtain 
\begin{subequations}
\begin{flalign}
\nabla_{\boldsymbol{\theta}^{l}}\bigl(\sum_{j=1}^{K}|\bigl(\boldsymbol{\theta}^{l}\bigr)\trans\mathbf{e}_{k,j}^{l}|^{2}+\sigma^{2}\bigr) & =\sum_{j=1}^{K}\mathbf{e}_{k,j}^{l\ast}\bigl(\mathbf{e}_{k,j}^{l}\bigr)\trans\boldsymbol{\theta}^{l},\label{eq:CF1}\\
\nabla_{\boldsymbol{\theta}^{l}}\bigl(\sum_{j\neq k}^{K}|\bigl(\boldsymbol{\theta}^{l}\bigr)\trans\mathbf{e}_{k,j}^{l}|^{2}+\sigma^{2}\bigr) & =\sum_{j\neq k}^{K}\mathbf{e}_{k,j}^{l\ast}\bigl(\mathbf{e}_{k,j}^{l}\bigr)\trans\boldsymbol{\theta}^{l}.\label{eq:CF2}
\end{flalign}
\end{subequations}
 Substituting (\ref{eq:CF1}) and (\ref{eq:CF2}) into (\ref{eq:gradftheta_l})
yields (\ref{eq:phi_derivative}).

\bibliographystyle{IEEEtran}
\bibliography{IEEEabrv,SIM_References} 

\begin{thebibliography}{10}
\providecommand{\url}[1]{#1}
\csname url@samestyle\endcsname
\providecommand{\newblock}{\relax}
\providecommand{\bibinfo}[2]{#2}
\providecommand{\BIBentrySTDinterwordspacing}{\spaceskip=0pt\relax}
\providecommand{\BIBentryALTinterwordstretchfactor}{4}
\providecommand{\BIBentryALTinterwordspacing}{\spaceskip=\fontdimen2\font plus
\BIBentryALTinterwordstretchfactor\fontdimen3\font minus
  \fontdimen4\font\relax}
\providecommand{\BIBforeignlanguage}[2]{{%
\expandafter\ifx\csname l@#1\endcsname\relax
\typeout{** WARNING: IEEEtran.bst: No hyphenation pattern has been}%
\typeout{** loaded for the language `#1'. Using the pattern for}%
\typeout{** the default language instead.}%
\else
\language=\csname l@#1\endcsname
\fi
#2}}
\providecommand{\BIBdecl}{\relax}
\BIBdecl

\bibitem{an2024stacked}
J.~An \emph{et~al.}, ``Stacked intelligent metasurface-aided {MIMO} transceiver
  design,'' \emph{{IEEE} Wireless Commun. Mag.}, vol.~31, no.~4, pp. 123--131,
  2024.

\bibitem{di2024state}
M.~Di~Renzo, ``State of the art on stacked intelligent metasurfaces
  communication, sensing and computing in the wave domain,'' in \emph{Eur.
  Conf. Antennas Propag. (EuCAP), Stockholm, Sweden}, 2025, pp. 1--3.

\bibitem{hassan2024efficient}
N.~U. Hassan \emph{et~al.}, ``Efficient beamforming and radiation pattern
  control using stacked intelligent metasurfaces,'' \emph{IEEE Open J. Commun.
  Soc.}, vol.~5, pp. 599--611, 2024.

\bibitem{an2023stacked}
J.~An \emph{et~al.}, ``Stacked intelligent metasurfaces for efficient
  holographic {MIMO} communications in {6G},'' \emph{{IEEE} J. Sel. Areas
  Commun.}, vol.~41, no.~8, pp. 2380--2396, 2023.

\bibitem{an2023stackedmu}
J.~An, M.~Di~Renzo, M.~Debbah, and C.~Yuen, ``Stacked intelligent metasurfaces
  for multiuser beamforming in the wave domain,'' in \emph{ICC 2023-IEEE Int.
  Conf. Commun.}\hskip 1em plus 0.5em minus 0.4em\relax IEEE, 2023, pp.
  2834--2839.

\bibitem{an2025stackedmu}
J.~An \emph{et~al.}, ``Stacked intelligent metasurfaces for multiuser downlink
  beamforming in the wave domain,'' \emph{{IEEE} Trans. Wireless Commun.},
  vol.~24, no.~7, pp. 5525--5538, 2025.

\bibitem{lin2024stacked}
S.~Lin \emph{et~al.}, ``Stacked intelligent metasurface enabled {LEO} satellite
  communications relying on statistical {CSI},'' vol.~13, no.~5, pp.
  1295--1299, 2024.

\bibitem{papazafeiropoulos2024achievablemu}
A.~Papazafeiropoulos \emph{et~al.}, ``Achievable rate optimization for large
  stacked intelligent metasurfaces based on statistical {CSI},'' vol.~13,
  no.~9, pp. 2337--2341, 2024.

\bibitem{hu2024joint}
Y.~Hu \emph{et~al.}, ``Joint beamforming and power allocation design for
  stacked intelligent metasurfaces-aided cell-free massive {MIMO} systems,''
  \emph{{IEEE} Trans. Veh. Technol.}, vol.~74, no.~3, pp. 5235--5240, 2025.

\bibitem{shi2024joint}
E.~Shi \emph{et~al.}, ``Joint {AP-UE} association and precoding for {SIM}-aided
  cell-free massive {MIMO} systems,'' \emph{{IEEE} Trans. Wireless Commun.},
  vol.~24, no.~6, pp. 5352--5367, 2025.

\bibitem{niu2024stacked}
H.~Niu \emph{et~al.}, ``Stacked intelligent metasurfaces for integrated sensing
  and communications,'' vol.~13, no.~10, pp. 2807--2811, 2024.

\bibitem{bahingayi2025scaling}
E.~E. Bahingayi, N.~S. Perovi{\'c}, and L.-N. Tran, ``Scaling achievable rates
  in {SIM}-aided {MIMO} systems with metasurface layers: A hybrid optimization
  framework,'' 2025, {Early Access}.

\bibitem{papazafeiropoulos2024achievable}
A.~Papazafeiropoulos \emph{et~al.}, ``Achievable rate optimization for stacked
  intelligent metasurface-assisted holographic {MIMO} communications,''
  \emph{{IEEE} Trans. Wireless Commun.}, vol.~23, no.~10, pp. 13\,173--13\,186,
  2024.

\bibitem{park2025sim}
E.~Park \emph{et~al.}, ``{SIM}-enabled hybrid digital-wave beamforming for
  fronthaul-constrained cell-free massive {MIMO} systems,'' \emph{arXiv
  preprint arXiv:2506.19090}, 2025.

\bibitem{perovic2024energy}
N.~S. Perovi{\'c}, E.~E. Bahingayi, and L.-N. Tran, ``Energy-efficient designs
  for {SIM}-based broadcast {MIMO} systems,'' \emph{arXiv preprint
  arXiv:2409.00628}, 2024.

\bibitem{shi2025energy}
E.~Shi \emph{et~al.}, ``Energy-efficient {SIM-assisted} communications: How
  many layers do we need?'' \emph{arXiv preprint arXiv:2504.15737}, 2025.

\bibitem{NemanjaSIM}
N.~S. Perovi{\'c} and L.-N. Tran, ``Mutual information optimization for
  {SIM}-based holographic {MIMO} systems,'' vol.~28, no.~11, pp. 2583--2587,
  2024.

\bibitem{Lin_2018}
X.~Lin \emph{et~al.}, ``All-optical machine learning using diffractive deep
  neural networks,'' \emph{Science}, vol. 361, no. 6406, pp. 1004--1008, Sep.
  2018.

\bibitem{shi2011iteratively}
Q.~Shi \emph{et~al.}, ``An iteratively weighted {MMSE} approach to distributed
  sum-utility maximization for a {MIMO} interfering broadcast channel,''
  \emph{{IEEE} Trans. Signal Process.}, vol.~59, no.~9, pp. 4331--4340, 2011.

\bibitem{bahingayi2024joint}
E.~E. Bahingayi, N.~S. Perovi{\'c}, and L.-N. Tran, ``On the joint beamforming
  design for large-scale downlink {RIS}-assisted multiuser {MIMO} systems,''
  \emph{arXiv preprint arXiv:2412.08320}, 2024.

\end{thebibliography}

\end{document}